\documentclass[aps,prl,reprint,preprintnumbers,superscriptaddress,amsmath,amssymb,floatfix,longbibliography]{revtex4-2}
\pdfoutput=1
\usepackage{subfigure}
\usepackage{graphicx}
\usepackage{amsmath}
\usepackage{bm}
\usepackage[colorlinks,citecolor=blue,urlcolor=blue,linkcolor=red]{hyperref}

\begin{document}
\title{Production Rate of Glueball-like $X(2370)$ in $J/\psi$
Radiative Decay}

\author{Ying Chen}
\email{cheny@ihep.ac.cn}
\affiliation{Institute of High Energy Physics, Chinese Academy of Sciences, Beijing 100049, People's Republic of China}
\affiliation{School of Physical Sciences, University of Chinese Academy of Sciences, Beijing 100049, People's Republic of China}
\author{Long-Cheng Gui }
\email{guilongcheng@hunnu.edu.cn}
\affiliation{Department of Physics, Hunan Normal University, and Key Laboratory of Low-Dimensional Quantum Structures and Quantum Control of Ministry of Education,   Changsha 410081, China}
\affiliation{Synergetic Innovation Center for Quantum Effects and Applications (SICQEA), Hunan Normal University, Changsha 410081, China }
\author{Geng Li}
\affiliation{Institute of High Energy Physics, Chinese Academy of Sciences, Beijing 100049, People's Republic of China}
\author{\small Wei Sun}
\affiliation{Institute of High Energy Physics, Chinese Academy of Sciences, Beijing 100049, People's Republic of China}



\begin{abstract}
$X(2370)$ falls in the mass region of the lowest pseudoscalar glueball predicted by lattice QCD studies and its decay properties are similar to those of $\eta_c$. 
A previous lattice QCD study finds that the pseudoscalar glueball ($G_{0^-}$) and the lowest pseudoscalar charmonium ($c\bar{c}(0^-)$) can mix with a small mixing angle $\sin\theta$. 
It is therefore possible that $X(2370)$ and $\eta_c$ are admixtures of $G_{0^-}$ and $c\bar{c}(0^-)$. 
In this picture, although $\mathrm{Br}(J/\psi\to\gamma\eta_c)$ is insensitive to the small $\sin\theta$, $\mathrm{Br}(J/\psi\to \gamma X(2370))$ can be enlarged drastically by the mixing due to the much larger kinematic factor for $J/\psi\to \gamma X(2370)$ and the much larger transition form factor for $J/\psi\to \gamma (c\bar{c}(0^{-}))$. 
Depending on the value of $\sin\theta$, $\mathrm{Br}(J/\psi\to \gamma X(2370))$ can be much larger than that of the pure pseudoscalar glueball, namely, $2.3(8)\times 10^{-4}$ that is predicted by a quenched lattice QCD calculation. 
Present results by BESIII favor a small mixing angle of $\mathcal{O}(1^\circ)$, which can be further constrained by more measurements of $X(2370)$ decays if the mixing picture applies here.       
\end{abstract}

\maketitle
\paragraph{Introduction---}
$X(2370)$ was first observed by the BESIII Collaboration (BESIII) in the invariant mass spectrum of $\pi^+\pi^-\eta'$ of the decay process $J/\psi\to \gamma \pi^+\pi^- \eta'$ in 2011~\cite{BESIII:2010gmv} and was confirmed by BESIII in the same process in 2016~\cite{BESIII:2016fbr}. 
$X(2370)$ was also observed in $J/\psi\to \gamma K\bar{K}\eta'$ by BESIII in 2019~\cite{BESIII:2019wkp}. 
Recently, BESIII performed a partial wave analysis of $J/\psi\to \gamma K_S^0K_S^0 \eta'$ and determined the quantum numbers of $X(2370)$ to be $J^{PC} = 0^{-+}$~\cite{BESIII:2023wfi}. 
Reference~\cite{Huang:2025pyv} also reports preliminary BESIII results showing the possible signals of $X(2370)$ in the $K_S^0K_S^0\eta$, $K_S^0 K_S^0\pi^0$, $\pi^0\pi^0\eta$ systems in $J/\psi$ radiative decays. 
These observations indicate that $X(2370)$ has a similar decay pattern to that of $\eta_c$. 
Since the decays of $\eta_c$ into light hadrons proceed through the annihilation of the charm quark and antiquark ($c\bar{c}$) into multiple (at least two) gluons, it is expected that $X(2370)$ also decays through gluonic intermediate states. 
On the other hand, lattice QCD studies, either in the quenched approximation~\cite{Morningstar:1999rf,Chen:2005mg,Gui:2019dtm,Athenodorou:2020ani} or with dynamical quarks~\cite{Sun:2017ipk,Chen:2021dvn,Athenodorou:2023ntf}, predict that the lowest pseudoscalar glueball ($G_{0^-}$) has a mass ranging from $2.3-2.7$~GeV. 
The mass of $X(2370)$ falls into this mass region and hints that $X(2370)$ could be a candidate for the pseudoscalar glueball. 

According to BESIII's observations, both $X(2370)$ and $\eta_c$ show up in the five three-pseudoscalar-meson systems mentioned above, and $X(2370)$ is an isolated structure in the energy region between 2.2~GeV and $m_{\eta_c}$ (except for the $\pi^+\pi^-\eta'$ spectrum that may accommodate a $X(2600)$ state~\cite{BESIII:2016fbr}). 
Since $X(2370)$ has a mass close to $m_{\eta_c}$ and exhibits a similar decay pattern to that of $\eta_c$, it is possible that $\eta_c$ and $X(2370)$ are mixed mass eigenstates of the lowest pure pseudoscalar $c\bar{c}$ state $(c\bar{c}(0^-))$ and the pseudoscalar glueball $G_{0^-}$. 
The $\eta-\eta'-\eta_c-G_{0^-}$ mixing was explored in phenomenological studies~\cite{Tsai:2011dp,Qin:2017qes}, which prefer a mixing angle about $11^\circ$. The $c\bar{c}(0^-)-G_{0^-}$ mixing was calculated also by a lattice QCD study with $N_f=2$ dynamical charm quarks~\cite{Zhang:2021xvl}. 
In this unitary lattice setup for charm quarks, the mixing energy $x=\langle (c\bar{c}(0^-))|H_I|G_{0^-}\rangle = 49(6)$~MeV is determined. 
With the approximations $m_{c\bar{c}(0^-)}\approx m_{\eta_c}$ and $m_{G(0^-)}\approx m_X$, the mixing angle is estimated to be $\sin\theta = 0.08(1)$, which implies that $\eta_c$ has a 99\% $c\bar{c}(0^-)$ component while $X(2370)$ is a $99\%$ pseudoscalar glueball. 
This mixing increases the total width $\Gamma_{\eta_c}$ of $\eta_c$ by roughly 7~MeV (using the total width $\Gamma_X\approx 100$~MeV of $X(2370)$) with respect to that of $c\bar{c}(0^-)$ and explains the relatively large $\Gamma_{\eta_c}=30.5(5)$~MeV~\cite{ParticleDataGroup:2024cfk} to some extent. 
It is seen that the small mixing angle can have sizeable effects for $\Gamma_X\gg \Gamma_{\eta_c}$. 
A similar effect of this mixing is expected in the production rate of $\eta_c$ and $X(2370)$ in the $J/\psi$ radiative decay since $\mathrm{Br}(J/\psi\to \gamma \eta_c)=1.41(14)\%$~\cite{ParticleDataGroup:2024cfk} is much larger than the lattice QCD prediction $\mathrm{Br}(J/\psi\to \gamma G_{0^-})=2.3(8)\times 10^{-4}$ when using $m_{G_{0^-}}\approx m_{X(2370)}$~\cite{Gui:2019dtm}. 
In this Letter, we will explore the production rate of $X(2370)$ in the $J/\psi$ radiative decay based on the $c\bar{c}(0^-)-G_{0^-}$ mixing.       

\paragraph{Formalism---}
For a pseudoscalar $P$, the partial decay width of $J/\psi\to \gamma P$ is related to the on-shell form factor $M^{(P)}(Q^2=0)$ as
\begin{equation}\label{eq:width}
    \Gamma(J/\psi\to \gamma P)=\frac{4\alpha}{27}|\vec{k}|^3 M^2(0),
\end{equation}
where the electric charge of charm quark $Q_c=2e/3$ has been incorporated, $\alpha=1/134$ is the fine structure constant at the charm quark mass scale, and $|\vec{k}|=(m_{J/\psi}^2-m_P^2)/(2m_{J/\psi})$ is the spatial momentum of the final state photon. 
The form factor $M(Q^2)$ is defined through the electromagnetic multipole decomposition~\cite{Dudek:2006ej} of the transition matrix element
\begin{equation}\label{eq:matrix}
     \langle P(k)|j^\mu(0)|J/\psi(p, \lambda)\rangle
    = M^{(P)}(Q^2) \epsilon^{\mu\nu\rho\sigma}p_{\nu}k_{\rho}\epsilon_\sigma^\lambda(p),
\end{equation}
where $Q^2=-(p-k)^2$ is the virtuality of the photon, $\epsilon_\sigma^\lambda(p)$ is the polarization vector of $J/\psi$ and $j^\mu=\bar{c}\gamma^\mu c$ is the electromagnetic current involving only the charm quark, since $P$ is either a $c\bar{c}$ state or a glueball.

\begin{figure}[t]
	\centering
    \includegraphics[width=0.8\linewidth]{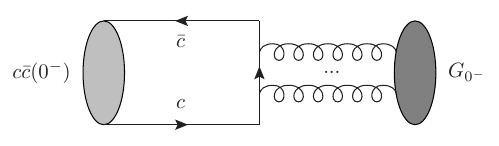}
	\caption{Schematic diagrams of $c\bar{c}(0^-)$--$G_{0^-}$ mixing. }
	\label{fig:topology-mix}
\end{figure}
Now we consider the mixing of the $c\bar{c}(0^-)$ state ($|c\bar{c}\rangle$) and the pure glueball $G_{0^-}$ state ($|G\rangle$). 
The mixing happens through the $c\bar{c}-gg(\cdots)$ dynamics (as shown in Fig.~\ref{fig:topology-mix}), where $gg(\cdots)$ denotes intermediate multi-gluon states (at least two gluons). If $X(2370)$ ($|X\rangle$) and $\eta_c$ ($|\eta_c\rangle$) are the corresponding two mass eigenstates after the mixing, one has 
\begin{equation}\label{eq:angle}
    \left(\begin{array}{c}
         |X\rangle   \\
         |\eta_c\rangle 
    \end{array}
    \right)=
    \left(\begin{array}{lr}
      \cos\theta &-\sin\theta \\
      \sin\theta & \cos\theta    
    \end{array}
    \right)
    \left(
    \begin{array}{c} |G\rangle \\ |c\bar{c}\rangle
    \end{array}
    \right),
\end{equation}
where $\theta$ is the mixing angle. 
Subsequently, the on-shell form factors $M^{(P)}(0)$ for $J/\psi\to \gamma X(\eta_c)$ processes are expressed as 
\begin{equation} \label{eq:mixm}
    \begin{aligned}
        M^{(X)}(0) &= M^{(G)}(0) \cos\theta-M^{(c\bar{c})}(0)\sin\theta,    \\
        M^{(\eta_c)}(0) &= M^{(G)}(0) \sin\theta+M^{(c\bar{c})}(0)\cos\theta.    
    \end{aligned}
\end{equation} 

\begin{figure}[t]
	\centering
    \subfigure[]{
        \includegraphics[width=0.66\linewidth]{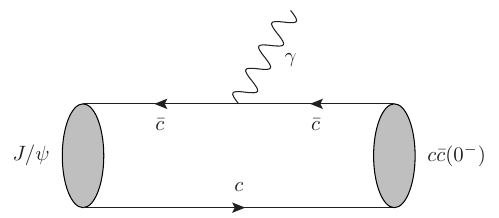}
        \label{fig:topology-decay-a}}
    \subfigure[]{
        \includegraphics[width=0.66\linewidth]{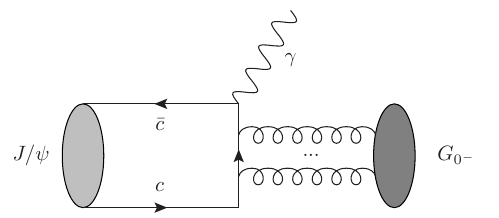}
        \label{fig:topology-decay-b}}
    \subfigure[]{
        \includegraphics[width=0.66\linewidth]{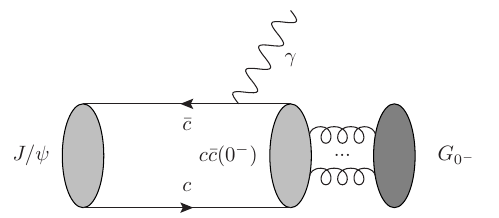}
        \label{fig:topology-decay-c}}
	\caption{Schematic diagrams for $J/\psi\to \gamma c\bar{c}(0^-)$ (diagram (a)) and $J/\psi\to\gamma G_{0^{-}}$ (diagram (b)). Diagram (c) depicts the mechanism that $J/\psi$ radiatively transitions into $c\bar{c}(0^-)$, which then mixes with $G_{0^-}$. }
	\label{fig:topology-decay}
\end{figure}

The on-shell form factors $M^{(c\bar{c})}(0)$ and $M^{(G)}(0)$ have been derived directly from the lattice QCD calculations. 
It should be noted that the lattice QCD calculations of $M^{(P)}(0)$ for $J/\psi\to \gamma \eta_c$ consider only the connected diagram shown in  Fig.~\ref{fig:topology-decay-a}, such that the obtained $M^{(P)}(0)$ is actually $M^{c\bar{c}}(0)$. The schematic diagram for the $J/\psi\to \gamma G_{0^-}$ is illustrated in Fig.~\ref{fig:topology-decay-b}. $M^{(P)}(0)$ is often expressed in terms of a dimensionless form factor $V^{(P)}(0)$ as 
\begin{equation}\label{eq:mtov}
    M^{(P)}(0)=\frac{2\omega_P V^{(P)}(0)}{m_{J/\psi}+m_P}.
\end{equation}
For $c\bar{c}(0^-)$, $\omega_P=2$ is to take into account the cases that the photon is emitted from both the charm quark and antiquark. 
For $G_{0^-}$, $\omega_P=1$ is taken since the $c\bar{c}$ annihilation produces a photon and gluons almost simultaneously with gluons coupling to $G_{0^-}$.    In practical lattice QCD calculations, the gluon lines in Fig.~\ref{fig:topology-decay-b} are realized by averaging over gauge configurations, while the photon is emitted from the charm quark loop. Therefore, all the possible sequences of the photon vertices and the gluon vertices are included implicitly, and there is no need to consider the position of the photon emission. 

Table~\ref{tab:ffv} collects the lattice QCD results of $V^{(c\bar{c})}(0)$, which are obtained either in the quenched approximation or with dynamical quarks, and are consistent with each other in general. Since the result by the HPQCD Collaboration~\cite{Colquhoun:2023zbc} is determined from the $N_f=2+1+1$ QCD (including the dynamical $u,d,s,c$ quark flavors) and systematic uncertainties are better controlled after the continuum and chiral extrapolations, we use the value $V^{(c\bar{c})}(0)=1.865(7)(8)$ in the following discussion. 
\begin{table}[t]
    \centering
    \caption{Values of the dimensionless form factor $V^{(c\bar{c})}(0)$ determined by quenched lattice QCD calculations (QA) and the calculations with dynamical light quarks ($N_f=2$ and $N_f=2+1$) till now.} 
    \label{tab:ffv}
    \begin{ruledtabular}
        \begin{tabular}{ll}
        Lattice setup & $V^{(c\bar{c})}(0)$\\
        \hline
        QA~\cite{Dudek:2006ej}                          & 1.85(4)\\
        QA~\cite{Gui:2019dtm}                           & 1.93(4)\\
        $N_f=2$~\cite{Chen:2011kpa}                     & 2.01(2)\\
        $N_f=2$~\cite{Becirevic:2012dc}                 & 1.92(3)(2)\\
        $N_f=2$~\cite{Li:2023zig}                                 & 2.08(1)\\
        $N_f=2+1$~\cite{Donald:2012ga}                  & 1.90(7)(1)\\
        $N_f=2+1$~\cite{Meng:2024axn}                   & 1.90(4)\\
        $N_f=2+1+1$~\cite{Colquhoun:2023zbc}            & 1.865(7)(8)\\
        \end{tabular}
    \end{ruledtabular}
\end{table}

There is only one lattice QCD calculation of $V^{(G)}(0)$ in the quenched approximation, which gives $V^{(G)}(0)=0.0246(43)$ and $m_{G_{0^-}}=2.395(14)~{\rm GeV}$ after the continuum extrapolation~\cite{Gui:2019dtm}. 
Then with the experimental mass values $m_{J/\psi}=3.0969(0)$~GeV and $m_{\eta_c}=2.9841(4)$~GeV~\cite{ParticleDataGroup:2024cfk}, as well as the value of $m_{G_{0^-}}$ mentioned above and the approximation $m_{c\bar{c}(0^-)}\approx m_{\eta_c}$, Eq.~\eqref{eq:mtov} gives 
\begin{equation} \label{eq:mvalue}
    \begin{aligned}
        M^{(c\bar{c})}(0)&=1.227(7)\,{\rm GeV}^{-1},\\
        M^{(G)}(0)&=0.0090(16)~\rm{GeV}^{-1}.
    \end{aligned}
\end{equation} 

The effect of the $c\bar{c}(0^-)-G_{0^-}$ mixing on the decays $J/\psi\to \gamma \eta_c(X)$ is illustrated in Fig.~\ref{fig:topology-decay-c}. 
We omit the other diagram describing the mixing of the final-state $G_{0^-}$ produced in $J/\psi\to\gamma G_{0^-}$ with $c\bar{c}(0^-)$ since $M^{(G)}(0)\ll M^{(c\bar{c})}(0)$. 
The difference between Fig.~\ref{fig:topology-decay-b} and Fig.~\ref{fig:topology-decay-c} is interpreted theoretically as follows: 
The former describes the $c\bar{c}$ pair in $J/\psi$ annihilating into a photon plus gluons that couple to $G_{0^-}$, while the latter means that after the photon emission, the $c\bar{c}$ pair develops a propagation of $c\bar{c}(0^-)$ that finally mixes with $G_{0^-}$. 
The mechanism in Fig.~\ref{fig:topology-decay-c} has not been investigated by the lattice QCD, which requires the numerically challenging calculation of the four-point functions like $\langle \mathcal{O}_{J/\psi}(t)j^\mu(t_1)\mathcal{O}_{c\bar{c}(0^-)}(t_2)\mathcal{O}_{G^{0^-}}(0)\rangle$, where $\mathcal{O}_Y$ refers to the lattice interpolation field for $Y=J/\psi, c\bar{c}(0^-), G_{0^-}$. 
Fortunately, the right parts of Figs.~\ref{fig:topology-decay-a},~\ref{fig:topology-decay-b},~\ref{fig:topology-decay-c} after the photon emission can be well described by the $c\bar{c}(0^-)-G_{0^-}$ mixing. 
It is commonly believed that $\eta_c$ is almost the lowest pseudoscalar charmonium, so the mixing angle $|\sin\theta|\ll 1$ is indispensable. 
Then combining Eqs.~(\ref{eq:width},~\ref{eq:mixm},~\ref{eq:mvalue}) and the experimental total width $\Gamma_{J/\psi}=92.6(1.4)~{\rm keV}$ of $J/\psi$~\cite{ParticleDataGroup:2024cfk}, one has 
\begin{equation} \label{eq:width-phy}
    \begin{aligned}
        \mathrm{Br}(J/\psi\to\gamma\eta_c)&= (2.44(5)\times 10^{-2})\cos^2\theta,\\
        \mathrm{Br}(J/\psi\to\gamma X)    &= 2.88(4)\times 10^{-4}\times \\
            &~~~\left(122.7(7)\sin\theta-0.90(16)\cos\theta\right)^2,
    \end{aligned}
\end{equation} 
where the $M^{(G)}\sin\theta$ term in $\Gamma(J/\psi\to \gamma\eta_c)$ is too tiny to be considered. 
Obviously, when $\cos\theta=1$ (no mixing), one recovers the lattice predictions in Ref.~\cite{Colquhoun:2023zbc} for $\Gamma(J/\psi\to \gamma\eta_c)$ and in Ref.~\cite{Gui:2019dtm} for $\Gamma(J/\psi\to \gamma G_{0^-})$.
When the mixing mechanism is considered, the mixing effect from $(c\bar{c}(0^-))$ on $\mathrm{Br}(J/\psi\to \gamma X(2370))$ is enlarged in two ways: 
First, Eq.~\eqref{eq:mvalue} indicates that $M^{(c\bar{c})}(0)\gg M^{(G)}(0)$. 
Secondly, the partial decay width is proportional to $|\vec{k}|^3$ with $\vec{k}$ being the photon momentum. 
With the experimental values of masses, one has 
\begin{equation}
    \frac{|\vec{k}(J/\psi\to \gamma X)|^3}{|\vec{k}(J/\psi\to \gamma \eta_c)|^3}=\left(\frac{m_{J/\psi}^2-m_{X}^2}{m_{J/\psi}^2-m_{\eta_c}^2}\right)^3\approx 177.5.
\end{equation}
Consequently, in contrast to $\Gamma(J/\psi\to \gamma\eta_c)$ that is insensitive to the mixing angle $\sin\theta$, the value of $\Gamma(J/\psi\to\gamma X(2370))$ has a very strong dependence on $\sin\theta$ even if it is small. 
In this sense, the value of $\sin\theta$ is crucial for the production rate of $X(2370)$. 

\begin{figure}[t]
	\centering
    \includegraphics[width=0.8\linewidth]{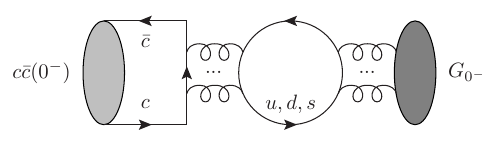}
	\caption{Schematic diagram for the contribution of light quarks to the $c\bar{c}(0^-)$--$G_{0^-}$ mixing.  }
	\label{fig:topology-light}
\end{figure}

Actually, a previous lattice QCD study explored the $c\bar{c}(0^-)-G_{0^-}$ mixing
using gauge ensembles with $N_f=2$ degenerate charm sea quarks~\cite{Zhang:2021xvl}. 
In this unitary lattice setup for charm quark, the mixing angle $\sin\theta$ is derived directly. 
At two $J/\psi$ masses, namely, $m_{J/\psi} = 2.743(1)$~GeV and $3.068(1)$~GeV, the mixing angle is determined to be $\theta = 6.6(9)^\circ$ and $4.3(4)^\circ$, respectively. Then based on the effective Hamiltonian
\begin{equation}
    H=\left(
    \begin{array}{cc}
         m_{G} & x\\
         x     &m_{c\bar{c}}         
    \end{array}
    \right),
\end{equation} 
the mixing energy $x= 49(6)$~MeV is estimated and is observed to be insensitive to the charm quark mass. 
This mixing energy may shift the masses $m_{G}$ and $m_{c\bar{c}}$ only by a few MeV, so the mixing angle is estimated to be  
\begin{equation}
    \sin\theta\approx \frac{x}{m_{\eta_c}-m_X}= 0.083_{-0.017}^{+0.011}, ~ \theta\approx (4.8_{-1.0}^{+0.6})^\circ,
\end{equation}
using the experimental values of $m_{\eta_c}$ and $m_X=2.395\pm 11_{-94}^{+26}$~GeV\cite{BESIII:2023wfi}.
Plugging this value into Eq.~\eqref{eq:width-phy}, one has 
\begin{equation} \label{eq:width-br}
    \begin{aligned}
    \Gamma(J/\psi\to\gamma X(2370))&\approx 2.3_{-1.0}^{+0.7}~{\rm keV},\\
    \mathrm{Br}(J/\psi\to \gamma X(2370))&\approx 2.5_{-1.1}^{+0.7}\times 10^{-2}.
    \end{aligned}
\end{equation} 
This branching fraction is obviously a huge value but is mainly due to the not too small mixing angle. 

\begin{table*}[t]
    \centering
    \caption{Combined branching ratios of $J/\psi\to \gamma X(2370)$ and $X\to PPP$ measured by BESIII. Here $PPP$ refers to three-pseudoscalar final states. } 
    \label{tab:BESIII}
    \begin{ruledtabular}
        \begin{tabular}{ll}
        Decay process                                    &   Combined branching fraction \\
        \hline
       $J/\psi\to \gamma X(2370)\to \gamma K^+K^-\eta'$  &  $(1.79\pm 0.23\pm 0.65)\times 10^{-5}$~\cite{BESIII:2019wkp}\\
       $J/\psi\to \gamma X(2370)\to \gamma K_S^0K_S^0\eta'$  &  $(1.18\pm 0.32\pm 0.39)\times 10^{-5}$~\cite{BESIII:2019wkp}\\
       $(J/\psi\to \gamma X(2370), X(2370)\to f_0(980)\eta',f_0(980)\to K_S^0K_S^0)$ & $(1.31\pm 0.22_{-0.84}^{+2.85})\times 10^{-5}$~\cite{BESIII:2023wfi}
        \end{tabular}
    \end{ruledtabular}
\end{table*}

We would like to remark that the mixing angle $\sin\theta\approx 0.08(1)$ is determined in a unphysical lattice setup with two flavors of charm sea quark and without light sea quarks. 
In this lattice setup, the flavor singlet pseudoscalar charmonium $(c\bar{c}(0^-))$ has the quark configuration $(c_1\bar{c}_1+c_2\bar{c}_2)/\sqrt{2}$ with $c_{1,2}$ referring to the two flavors of degenerate charm quark. 
Since $c_1\bar{c}_1$ and $c_2\bar{c}_2$ have the same coupling mechanism to gluons, as illustrated in Fig.~\ref{fig:topology-mix}, the value of $\sin\theta$ mentioned above can be matched qualitatively to the one flavor charm quark case by a factor of $1/\sqrt{2}$, which will reduce the branching fraction in Eq.~\eqref{eq:width-br} by one half. 
On the other hand, when the light sea quarks are present, their contributions to the $c\bar{c}(0^-)-G_{0^-}$ mixing should be via the light quark loops, as shown in Fig.~\ref{fig:topology-light}. 
In the flavor singlet pseudoscalar channel, the coupling of a quark-anti-quark pair to gluons can be nonperturbative and can be enhanced due to the QCD $\mathrm{U}_A(1)$ anomaly that introduces a topological $q\bar{q}-gg(\cdots)$ coupling~\cite{Bass:2018xmz}. 
This argument is supported by the experimental observations that light pseudoscalars (such as $\eta'$ and $\eta(1405)$) usually have large production rates in the $J/\psi$ radiative decay~\cite{ParticleDataGroup:2024cfk}. 
Based on the $U_A(1)$ anomaly, a $N_f=2$ lattice QCD study gives a value of $\mathrm{Br}(J/\psi\to \gamma \eta')$ that is consistent with the experimental value~\cite{Jiang:2022gnd}. 
So the light quark loop contribution can evade the OZI suppression that usually applies to other channels and can result in sizable effect on the $c\bar{c}(0^-)-G_{0^-}$ mixing. 
Comparing Fig.~\ref{fig:topology-mix} and Fig.~\ref{fig:topology-light}, the latter has an additional (light) quark loop that introduces a relative minus sign to Fig.~\ref{fig:topology-mix}. 
Therefore, the light quark contribution reduces, to some extent, the $c\bar{c}(0^-)-G_{0^-}$ mixing through the dynamics of Fig.~\ref{fig:topology-mix}. 
This minus sign is also observed in the lattice QCD calculation of the form factors for the $D_s\to \eta(\eta')$ semileptonic decay~\cite{Bali:2014pva}. 
Based on the discussion above, the branching fraction of $J/\psi\to \gamma X(2370)$ in the real world is likely smaller than the value in Eq.~\eqref{eq:width-br}.   

\begin{figure}[t]
	\centering
    \includegraphics[width=\linewidth]{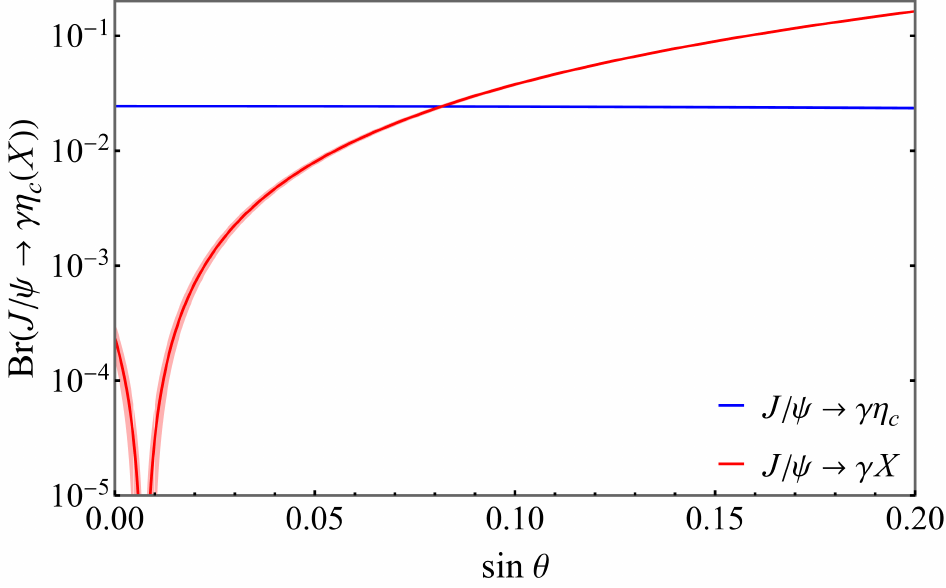}
	\caption{The branching ratios (Br) of $J/\psi \to \gamma \eta_c (X)$ processes as functions of the mixing angle $\theta$. The shaded bands represent theoretical uncertainties. }
	\label{fig:br}
\end{figure}

Nevertheless, the study in Ref.~\cite{Zhang:2021xvl} gives strong evidence from first principles that $c\bar{c}(0^-)$ and $G_{0^-}$ do mix, although the derived mixing angle has uncontrolled systematic uncertainties. 
So the result in Eq.~\eqref{eq:width-phy} is robust. 
Since the precise value of the mixing angle $\sin\theta$ has not been finally determined, we plot in Fig.~\ref{fig:br} the branching ratios of the processes $J/\psi\to \gamma\eta_c,X(2370)$ with respect to $\sin\theta$ in the range $\sin\theta\in [0,0.2]$. 
More experimental results of $\mathrm{Br}(J/\psi\to \gamma X(2370)\to \gamma+{\rm (light~hadrons)}$ are needed to constrain the value of $\sin\theta$.  

\paragraph{Discussion---} 
First, we discuss the experimental result of $\mathrm{Br}(J/\psi\to\gamma\eta_c)$. 
The world average value is $1.41(14)\%$ by PDG 2024~\cite{ParticleDataGroup:2024cfk}, which shows a tension with the lattice results $2.44(5)\%$ (see Eq.~\eqref{eq:width-phy} for $\cos\theta=1$). 
Recently, BESIII reported the latest experimental result $(2.29\pm0.01_{\rm stat}\pm 0.04_{\rm syst}\pm 0.18_{\rm opbf}\%)$ (Here `opbf' refers to the uncertainty from the other product branching fractions) from the process $(J/\psi\to \gamma\eta_c, \eta_c\to p\bar{p})$~\cite{BESIII:2025vdn}, which is consistent with the lattice QCD result. 
This value favors $\cos\theta\approx 1$ and supports that $\eta_c$ is predominantly a pure charmonium if the $c\bar{c}(0^-)-G_{0^-}$ mixing applies here. 

Now we switch to the production rate of $X(2370)$ in the $J/\psi$ radiative decay. 
To date, BESIII has reported three combined branching fractions, as listed in Table~\ref{tab:BESIII}. 
According to these branching fractions, the SU(3) flavor symmetry implies $\mathrm{Br}(J/\psi\to\gamma X(2370)\to \gamma K\bar{K}\eta')\sim 4\times 10^{-5}$. 
If $X(2370)$ is the pure pseudoscalar glueball, the quenched lattice QCD study predicts $\mathrm{Br}(J/\psi\to \gamma X(2370))=2.3(8)\times 10^{-4}$~\cite{Gui:2019dtm}. Then it seems that $K\bar{K}\eta'$ is a major decay mode of a branching fraction about 20\%. 
However, given a glueball state, the light hadron decays of $X(2370)$ should have a similar pattern to that of $\eta_c$, since these decays are through intermediate gluons. 
The branching fraction of $\eta_c\to K\bar{K}\eta'$ is only $1.73(35)\%$, so $\mathrm{Br}(X(2370)\to K\bar{K}\eta')\sim 20\%$ is implausibly large. 
The most probable reason for this inconsistency is that the production rate of $X(2370)$ in the $J/\psi$ radiative decay is much larger than that of a pure pseudoscalar glueball (This issue was also addressed in a phenomenological study~\cite{Sun:2021kka}).  

The $c\bar{c}(0^-)-G_{0^-}$ mixing helps understand the large production rate of $X(2370)$, as shown in Eq.~\eqref{eq:width-phy} and Fig.~\ref{fig:br}. 
In other words, a small mixing angle $\sin\theta$ can lead to a large production rate. 
The preliminary results of BESIII~\cite{Huang:2025pyv} indicate that, $X(2370)$ and $\eta_c$ appear simultaneously in the five three-pseudoscalar systems in the $J/\psi$ radiative decay, namely, $K_S^0K_S^0\eta, K_S^0K_S^0\pi^0, \pi^0\pi^0\eta$ as well as $\pi^+\pi^-\eta'$ and $K_S^0K_S^0\eta'$. 
Especially, there is one clear structure (likely $X(2370)$) appearing in the energy region from 2.2~GeV to $m_{\eta_c}$~\cite{BESIII:2019wkp,Huang:2025pyv} except for the $\pi^+\pi^-\eta'$ system~\cite{BESIII:2010gmv,BESIII:2016fbr} where an additional structure $X(2600)$ was observed. 
This indeed shows the similarity of $X(2370)$ and $\eta_c$ decays and is consistent with the mixing mechanism. 
If the branching fraction $\mathrm{Br}(X(2370)\to K\bar{K}\eta')$ is also roughly 2\% like that of $\eta_c$ ($\mathrm{Br}(\eta_c\to K\bar{K}\eta')=1.73(35)\%$~\cite{ParticleDataGroup:2024cfk}), the estimated $\mathrm{Br}(J/\psi\to\gamma X(2370)\to \gamma K\bar{K}\eta')\sim 4\times 10^{-5}$ from Table~\ref{tab:BESIII} gives $\mathrm{Br}(J/\psi\to X(2370))\sim 2\times 10^{-3}$. 
Then the mixing angle is estimated to be $\sin\theta \sim 0.03$ or $\theta\sim 2^{\circ}$ using Eq.~\eqref{eq:width-phy}. 
This mixing angle is roughly three times smaller than $\sin\theta =0.08(1)$ or $\theta=4.6(6)^\circ$ in Ref.~\cite{Zhang:2021xvl} and the reasons have been discussed above. 
This small mixing angle indicates that $\eta_c$ is mostly a pure pseudoscalar charmonium, and $X(2370)$ is almost a pure pseudoscalar glueball in the $c\bar{c}(0^-)-G_{0^-}$ mixing picture while it production in the $J/\psi$ radiative decay is mainly through its tiny $c\bar{c}(0^-)$ component.   

\paragraph{Summary---} 
The BESIII Collaboration observed the simultaneous appearance of $\eta_c$ and the pseudoscalar meson $X(2370)$ in the three-pseudoscalar systems $\pi^+\pi^-\eta'$~\cite{BESIII:2010gmv,BESIII:2016fbr}, $K\bar{K}\eta'$~\cite{BESIII:2019wkp,BESIII:2023wfi} (and likely $K_S^0K_S^0\eta$, $K_S^0K_S^0\pi^0$ and $\pi^0\pi^0\eta$~\cite{Huang:2025pyv}) in the $J/\psi$ radiative decay. 
Especially, $X(2370)$ appears to be the only structure in the energy region from 2.2~GeV to $m_{\eta_c}$, which accommodates the pseudoscalar glueball predicted by lattice QCD studies. 
On the other hand, a previous lattice QCD study found that the pseudoscalar glueball ($G_{0^-}$) and the lowest pseudoscalar charmonium ($c\bar{c}(0^-)$) can mix with a small mixing angle $\sin\theta$. 
So there is a possibility that $X(2370)$ and $\eta_c$ are admixtures of $G_{0^-}$ and $c\bar{c}(0^-)$. 
Within the mixing picture, we discuss the production rate of $X(2370)$ in the $J/\psi$ radiative decay. 
It is found that, although $\mathrm{Br}(J/\psi\to\gamma\eta_c)$ is insensitive to the small $\sin\theta$, $\mathrm{Br}(J/\psi\to \gamma X(2370))$ can be enhanced drastically by the mixing due to the much larger kinematic factor for $J/\psi\to \gamma X(2370)$ and the much larger transition form factor for $J/\psi\to \gamma (c\bar{c}(0^{-}))$. In other words, depending on the value of $\sin\theta$, the production of $X(2370)$ in the $J/\psi$ radiative decay is mainly through its tiny $c\bar{c}(0^-)$ component and can have a much larger production rate than $2.3(8)\times 10^{-4}$ of a pure pseudoscalar glueball. 
Present results by BESIII favor a small mixing angle of $\mathcal{O}(1^\circ)$. This implies that, in the $c\bar{c}(0^-)-G_{0^-}$ mixing picture, $\eta_c$ is mostly a pure pseudoscalar charmonium and $X(2370)$ is almost a pure pseudoscalar glueball. More experimental measurements of the decay modes of $X(2370)$ are desired to constrain the mixing angle.

\begin{acknowledgments}
\paragraph{Acknowledgments.---}
This work is supported by the National Natural Science Foundation of China (NNSFC) under Grants No.~12293060, No.~12293065, No.~12175063. 
WS and GL are also supported by Chinese Academy of Sciences under Grant No.~YSBR-101. 
GL is also supported by the China Postdoctoral Science Foundation under Grant No.~2025M773362.   
\end{acknowledgments}

\bibliography{ref}

\clearpage

\end{document}